\documentclass[useAMS,usenatbib]{mn2e}

\title[Spectral properties of Fermi blazars]
  {An Investigation into the Spectral Properties of Bright \textit{Fermi} Blazars}
\author[Harris et al.]
  {J.~Harris,$^1$\thanks{E-mail:j.d.harris@durham.ac.uk}
  P.M.~Chadwick$^1$ and
  M.K.~Daniel$^{1,2}$ \\
  $^1$Department of Physics, Durham University, South Road, Durham DH1 3LE, UK\\
  $^2$University of Liverpool, Oliver Lodge Laboratory, Liverpool L69 7ZE, UK}

\date{Released 2002 Xxxxx XX}

\pagerange{\pageref{firstpage}--\pageref{lastpage}} \pubyear{2002}

\def\LaTeX{L\kern-.36em\raise.3ex\hbox{a}\kern-.15em
    T\kern-.1667em\lower.7ex\hbox{E}\kern-.125emX}

\usepackage{graphicx}

\begin{document}

\label{firstpage}

\maketitle

\begin{abstract}
We investigate the spectral properties of blazars detected with the \textit{Fermi}-Large Area Telescope (LAT) in the high energy regime $100\rm~MeV$~-~$100\rm~GeV$.  We find that over long timescales a log-parabola provides an adequate description of the spectrum in almost all objects and in most cases is significantly better than a simple power law or broken power law description.  Broken power law descriptions appear to arise from two causes: confusion with nearby sources and as an artifact of older LAT instrument response functions.

We create a light curve for 2FGLJ2253.9+1609 (3C~454.3), the brightest of the objects investigated. During the quiescent state we find the spectrum to be fairly stable and well-described by a log-parabola.  There is some evidence that, on average, the peak energy of the inverse Compton emission is lower in the quiescent state than in the time-averaged state, suggesting that increases in flux are due to changing parameters within the jet as opposed to changes in an external photon field.  However, no correlation between inverse Compton peak energy and flux is apparent.  During high flux states, deviation of the spectral shape from a simple power law continues.  In some cases a log-parabola provides a significantly better fit than a broken power law but in others the reverse is true.
\end{abstract}


\section{Introduction}

The commonly accepted paradigm is that Active Galactic Nuclei (AGN) consist of a central supermassive black hole fed by an accretion disc.  The black hole is surrounded by two regions of material: the narrow line region (NLR) and broad line region (BLR) which are in turn surrounded by a dusty torus.  Some AGN have jets flowing at relativistic speeds and aligned roughly perpendicular to the dusty torus.  If the object is orientated such that its jet axis is at a small angle relative to the observer's line-of-sight it is termed a blazar \citep{Urry95}.

Blazars can be subdivided into two categories: BL Lacertae objects (BL Lacs) and Flat Spectrum Radio Quasars (FSRQs).  In BL Lacs, emission lines from the BLR are weak or not observed.

Blazars can emit radiation across the whole electromagnetic spectrum.  The spectral energy distribution (SED) of a blazar contains two broad peaks when plotted in units of $\log E$ against $\log EF(E)$, where $E$ is the energy of the radiation and $F(E)$ is the flux at that energy.  The low-energy peak occurs between infrared and X-ray frequencies and is attributed to synchrotron emission from leptons in the jet \citep{Fossati98}.  The high-energy peak, where observed, usually occurs at $\gamma $-ray frequencies.  The $\gamma $-ray emission is thought to originate in the relativistic jet, but the location of the emission within the jet and the mechanism for producing such high energies are uncertain. The emission is thought to be due to Compton upscattering of low energy seed photons.  These seed photons may be the synchrotron emission, known as the synchrotron self Compton (SSC) scenario \citep{Maraschi92}.   Alternatively they may originate from outside the jet, known as the external Compton (EC) scenario \citep{Dermer93} for example from the BLR or dusty torus \citep{Sokolov05}.  EC emission is thought to play a more prominent role in FSRQs due to the abundance of seed photons from the BLR.  This picture is further supported by the fact that the high-energy peak in FSRQs generally occurs at a lower energy than in BL Lacs, suggesting that the cooling mechanism is more efficient \citep{Ghisellini09}.

The \textit{Fermi}-Large Area Telescope (LAT), a space-based $\gamma $-ray telescope sensitive between $\sim 30\rm~MeV$ and $\sim 300\rm~GeV$, has detected several hundred blazars \citep{Nolan12}. Although roughly equal numbers of FSRQs and BL Lacs have been detected, the FSRQs tend to have higher-significance detections, since their high-energy peak occurs in a range where \textit{Fermi} has greater sensitivity.

As well as shedding light on the physics of blazars, a detailed description of the SED is necessary to provide improved constraints for the modelling of sources, which involves fitting to multiwavelength data.  It is also important to characterise the shape of the SED well in order to extrapolate up to $\rm TeV$ energies where the intrinsic spectrum may suffer pair absorption with extragalactic background light photons (see e.g., Orr, Krennrich and Dwek 2011).

In an earlier paper (Harris, Daniel \& Chadwick 2012, hereafter Paper~I) we investigated the spectral shape of $9$ \textit{Fermi}-bright FSRQs to determine the presence of curvature or breaks. All of the objects showed significant deviation from a simple power law, with $4$ of the objects being best described by a curved spectrum and $5$ by a spectrum with a sharp break.  For $5$ of the objects the spectral shape displayed significant variation with time.  \citet{Poutanen10} have demonstrated that if the emission region is inside the BLR, the spectrum would acquire a consistent break at $4$-$7\rm~GeV$ in the object rest frame due to pair absorption with photons emitted by Helium recombination.  However, since some objects examined in Paper~I lacked significant breaks and others had breaks that varied with time or were inconsistent with a $4$-$7\rm~GeV$ break, the model of \citet{Poutanen10} was strongly disfavoured.

In this work we have extended our sample to include more objects of both the FSRQ and BL Lac classes and extended the time period used from $2.5$ years up to $3.9$ years.  We have also examined the effects nearby confusing sources have on the results and investigated the behaviour in high flux states. We tested three spectral forms to determine which described the data from each object the best.  The first spectral form tested was the simple power law (SPL),

\begin{equation}\label{simplepl}
\frac{dN}{dE}=k\left(\frac{E}{E_0}\right)^{-\Gamma},
\end{equation}
 where ${dN}$/${dE}$ is the differential photon flux as a function of photon energy, $k$ is a normalisation constant, $E_0$ is a normalisation energy, which we fixed at $1\rm~GeV$, and $\Gamma$ is the spectral index. The second spectral form tested was a broken power law, which describes a sudden change in the spectral index of a power law from $\Gamma_1$ to $\Gamma_2$ at a break energy $E_b$,

\begin{equation}\label{brokenpl}
{\frac{dN}{dE}}= \left\{
\begin{array}{lll}
 k\left(\frac{E}{E_b}\right)^{-\Gamma_1}	&	{\rm if}	& E<E_b\\
 k\left(\frac{E}{E_b}\right)^{-\Gamma_2}&	{\rm if}	& E\geq E_b
\end{array}
\right\}.	
\end{equation}
A sharp break such as this could indicate multiple sources of seed photons, with the emission above the break due to scattering that occurs in the Klein-Nishina regime \citep{Finke10}.  The final spectral form tested was the log-parabola (LP) which describes constant curvature across the spectrum,
\begin{equation}\label{logparab}
\frac{dN}{dE}=k\left(\frac{E}{E_0}\right)^{-\Gamma -\beta\ln(\frac{E}{E_0})},
\end{equation}
where $\beta$ defines the amount of curvature and $E_0$ is again fixed at $1\rm~GeV$.  As shown by \citet{Massaro04}, spectra well described by an LP arise if leptons in the jet undergo stochastic acceleration processes.  A log-parabolic spectrum naturally has a peak energy, so if the peak of a blazar's SED occurs in the energy range being examined, a log-parabola would be expected to be the best description. In this case, determining that an LP was a better description than a BPL would not reveal much interesting information.  Therefore, as we describe later, we sometimes considered the SED only over a truncated energy range away from the peak energy such that if any spectral breaks are present in the data they would have a significant impact on model fitting.

We also performed a light curve analysis of the brightest object in our sample, 2FGLJ2253.9+1609 (3C~454.3), to identify high flux states.  The spectral shapes of these high states were then analysed separately.

The paper is structured as follows.  Section~$2$ gives a description of the selection cuts used in the \textit{Fermi} analysis, the sample selection, Monte Carlo simulations to identify objects with a high potential for source confusion, the method of searching for breaks and curvature in the SED and our light curve analysis.  Our results are presented and discussed in Section~$3$ and our conclusions are given in Section~$4$.

\section{Methods}

\subsection{Event selection}

For each target we examined all photons reconstructed to originate from a $15^{\circ}$ radius region of interest around the position of the source with an energy between $100\rm~MeV$ and $100\rm~GeV$ and detected between MJD~54643 and MJD~56085.  These were analysed using the \textit{Fermi} Science Tools v9r27p1 and the P7SOURCE\_V6 version of the \textit{Fermi}-LAT instrument response function (IRF) \citep{Ackermann12}.  Following standard practice,\footnote{
http://fermi.gsfc.nasa.gov/ssc/data/analysis/scitools/likelihood\_tutorial.html} we required that events for analysis had been suitably reconstructed (event class 2 (diffuse) or better) and discarded any photons which arrived while \textit{Fermi}'s zenith angle was $<100^{\circ}$ to avoid $\gamma-$ray contamination from the Earth.

Throughout the paper, the isotropic and Galactic $\gamma $-ray emission were fitted using the iso\_p7v6source and gal\_2yearp7v6 models respectively\footnote{http://fermi.gsfc.nasa.gov/ssc/data/access/lat/BackgroundModels.html} with the normalisations left as free parameters.

\subsection{Sample selection}

We initially selected all objects in the \textit{Fermi} 2-Year Point Source Catalog (2FGL, Nolan et al. 2012) which were identified as a BL Lac or FSRQ, had a test statistic (TS) of $2500$ or more and were at Galactic latitude of $10^{\circ}$ or more (to avoid contamination from the Milky Way).  This gave a sample of $15$~BL Lacs and $27$~FSRQs.

The point spread function of the \textit{Fermi}-LAT is as large as $3^{\circ}$ at $100\rm~MeV$ \citep{Nolan12} and it was necessary to ensure that the spectra of our target sources could be correctly identified, given the potential source confusion.  To this end we ran Monte Carlo simulations using the standard tool \textit{gtobssim}.  For each of our target sources, we created an input model which included the target source and all point sources in the 2FGL within $3^{\circ}$.  The target source was modelled as an SPL with the best fit values from the 2FGL. The confusing sources were each modelled as an SPL with values selected randomly from a normal distribution about their 2FGL values.

This input model was then used to generate simulated events equivalent to an observation spanning MJD~54643 to MJD~56085.  This simulated observation was then analysed using a binned analysis in \textit{gtlike} with the parameters of the target source left free and the parameters of the confusing sources fixed to their 2FGL values.

This method was repeated $100$ times to determine how often the input value of the target source's spectral index was included in the $68\%$ confidence interval (`success') or not.  If no source confusion occurs, success should occur in $68\%$ of cases with some random error.  Therefore, the mean success rate of target sources which do not suffer confusion should be $68\%$.  Confusion from nearby sources could systematically bias individual target sources such that the success rate was less than $68\%$.  Therefore we repeatedly removed the target source with the lowest success rate until the mean success rate was $68\%$.  The remaining sources constituted our `clean' sample while those that had been removed constituted our `unclean' sample.  Our clean sample consisted of $5$~BL Lacs and $15$~FSRQs and our unclean sample consisted of $10$
BL Lacs and $12$~FSRQs.  These samples are listed in Tables~\ref{AICTable} and~\ref{AICTableContam}.

\subsection{Searching for breaks and curvature \label{sec:sbc}}

As mentioned in Section~1, it was necessary to determine a suitable, truncated,  energy range, away from the peak in the SED, over which to search for spectral features.  Further discussion on this subject and validation of the following method for determining the energy range are given in Paper~I.

A binned analysis of the region of interest was performed and then the source was modelled as an LP with free parameters.  The peak energy, $E_{p}$, which corresponds to the high-energy peak of the SED, $\log E_{P}F(E_{P})$, was then estimated as
\begin{equation}\label{PeakEn}
E_{p}=E_{0} e^{\left(2-\Gamma\right)/2\beta},
\end{equation}
(see, e.g. Massaro 2004, but note that the logarithms in this paper are being taken to base $e$ rather than base $10$). If $E_{p}$ was estimated to occur below our maximum energy of $100\rm~GeV$ we classed it as having a low spectral peak and otherwise classed it as having a high spectral peak. The spectrum was then fitted with a BPL.  For a low (high) spectral peak, the minimum (maximum) energy was then increased (decreased) by $100\rm~MeV$ and a BPL fitted again.  If the break energies of the two BPLs were not consistent with one another to the $68\%$ confidence limit the process was repeated until this condition was satisfied.  For the majority of sources a consistent break was found after the first truncation in energy and therefore the whole energy range was used to find spectral parameters.

Once an energy range returning a stable break was found an SPL, BPL and LP were all fitted in this energy range. An Akaike Information Criterion ($AIC$) test was then performed to determine which, if any, was significantly better at describing the data.  The $AIC$ of a model $s$ (where $s$ models the source as SPL, BPL or LP) is given by
\begin{equation}
AIC_{s}=-2\ln L_{s}+2k_{fs},
\end{equation}
where $L_{s}$ is the likelihood of the model given the data and $k_{fs}$ is the number of free parameters in the model.  In model comparison, the difference between the $AIC$ of two models, $s$ and $s'$,
\begin{equation}\label{deltaAIC}
\Delta AIC_{s',s}=AIC_{s}-AIC_{s'},
\end{equation}
estimates the relative Kullback-Leibler information quantities of the two models \citep{Burnham01}, i.e. how much more model $s$ diverges from the true distribution of the data than model $s'$, or alternatively how much more information is lost by describing the data using $s$ rather than $s'$.  This is true for both nested and non-nested models (like BPL and LP) \citep{Findley88}.  The $AIC$ balances the systematic error in a model with fewer parameters with the random error in a model with more \citep{Bozdogan87}. A lower $AIC$ means a better description of the data, so when comparing multiple models it is useful to examine the difference in $AIC$ between a model $s$ and the minimum $AIC$ found, $AIC_{min}$,
\begin{equation}\label{deltaAICmin}
\Delta AIC_{min,s}=AIC_{s}-AIC_{min}.
\end{equation}
The best model by definition has $\Delta AIC_{min,s}=0$ and any model with $\Delta AIC_{min,s}>2$ is generally considered significantly worse \citep{Lewis11}. (Equation~\ref{deltaAIC} can be compared to the test statistic of a likelihood ratio  $-2\ln (L_{a}/L_{b})=-2(\ln L_{a} - \ln L_{b})$ which approximately follows a $\chi^{2}$ distribution.)

\subsection{Light curve analysis}
Making a light curve necessarily splits the dataset from an object and therefore we did this only for the brightest object in our sample.  We created a daily light curve for FSRQ 2FGLJ2253.9+1609 (3C~454.3) in the energy range $100 \rm~MeV$ to $100\rm~GeV$. The daily time bins were then grouped into blocks of consistent flux using the Bayesian blocks technique \citep{Scargle13}.  Identifying blocks of consistent flux is important because different emission mechanisms may be in effect during e.g. the stages of rising and falling flux.  Blocks identified as being in a high flux state were then examined to see if they were best described by an SPL, BPL or LP. We defined high flux as $>5\times 10^{-6}\rm~ph~cm^{-2}~s^{-1}$. This threshold was required to distinguish between BPL and LP spectra on daily timescales. The method for analysing each block was the same as described in Section~\ref{sec:sbc} but using an unbinned \textit{Fermi} analysis, which was computationally feasible on these smaller datasets.  We also analysed together all of the data from the quiescent blocks (which we defined as having flux $\le$ the average flux plus one standard deviation) and compared the result to the high flux blocks.

\section{Results and Discussion}
\subsection{Searching for breaks and curvature}

The results of searching for breaks and curvature are shown in Tables~\ref{AICTable} and~\ref{AICTableContam}.  Of the $5$ BL Lacs in the clean sample, $3$ were described significantly better by an LP and $2$ showed significant deviation from an SPL, but in these cases distinguishing between a BPL and an LP was not possible. Of the $15$~FSRQs in the clean sample, $11$ were described significantly better by an LP, $1$ object was best described by an SPL and $3$ showed significant deviation from an SPL but distinguishing between a BPL and an LP was not possible.
The single object in our clean sample that was best described by an SPL was 2FGLJ0957.7+5522 (4C~+55.17) which was found to have a spectrum of $\frac{dN}{dE}=\left(1.9\pm0.4\right) \times 10^{-11} \left(\frac{E}{10^{3}\rm~MeV}\right)^{-2.33\pm 0.09} \rm~ph~cm^{-2}~s^{-1}~MeV^{-1}$.

In the clean sample, more than half the objects were described significantly better by an LP than an SPL or a BPL, and all the sources are consistent with the hypothesis of an LP spectrum (i.e. neither an SPL or a BPL were a significantly better description than an LP).  Given that LP fits seemed ubiquitous and this spectral shape is capable of describing the area around the peak energy of the SED, we reran the analysis on the clean sample using an LP description over the entire, untruncated, energy range.  The results of these fits are shown in Table~\ref{LPTable}.  Among the $10$ objects that required the use of a truncated energy range in the earlier analysis, $8$ were consistent at the $1\sigma$ confidence level in the value of $\Gamma$ found using the whole energy range and $4$ were consistent in the value of $\beta$.  Among the values that did not agree there was roughly equal scatter in the values found using the whole energy range above and below the values found using the truncated energy range.  This suggests that no bias is introduced in the values for $\Gamma$ and $\beta$ by truncating the energy range, although using the whole energy range of course includes more data and is therefore preferred.  Both BL Lacs and FSRQs show a wide range of values for $\Gamma$ and $\beta$.  However, T-tests determined that (as might be expected from Section~1), compared to BL Lacs, FSRQs have significantly softer spectral indices ($\Gamma$), smaller curvature ($\beta$) and lower Compton peak energies ($E_{P}$) with p-values \footnote{The probability due to chance alone of observing a difference between the samples at least as extreme as the one found.} of $0.002$, $0.01$ and $0.04$ respectively.  With a p-value of $0.41$, the difference in the maximum value of the SED, $\log E_{P}F(E_{P}$), found for BL Lacs and FSRQs in the clean sample was not significant, which is most likely a selection bias where only objects above a certain brightness are included.

Of the $10$ BL Lacs in our unclean sample, $3$ were described significantly better by an LP, $2$ were described significantly better by a BPL, $4$ showed significant deviation from an SPL but distinguishing between a BPL and an LP was not possible, and $1$ object showed no significant deviation from an SPL.  Of the $12$ FSRQs in our unclean sample, $7$ were described significantly better by an LP, $1$ was described significantly better by a BPL, and $4$ showed significant deviation from an SPL but distinguishing between a BPL and an LP was not possible.

The unclean sample contained the only sources that were identified as having BPL spectra, and a higher incidence of sources where BPL and LP fits could not be distinguished from one another.  Caution should be exercised when working with sample sizes this small, but it seems probable that BPL fits can be favoured as a result of source confusion.  The BPL model has one more free parameter than the LP model, and so it would not be expected that a BPL would be found to be a better description of fainter objects simply because the photon numbers are smaller, in fact the opposite is true.

The results found in this work were compared to those in the 2FGL.  In the 2FGL, BPL fits were not considered and a likelihood ratio of $8$ was required to classify an object's spectrum type as LP, which corresponds roughly to $\Delta AIC_{LP,SPL} > 18$.  There were two objects, J0818.2+4223 and  J1159.5+2914, which were reported in the 2FGL as having SPL spectra but which we found to have $\Delta AIC_{LP,SPL} > 18$. As a check of the methods used in this paper, we re-ran our analysis of these two objects using the smaller time cuts from the 2FGL.  Both objects were found to have spectra consistent with the SPL reported in the 2FGL, with $\Delta AIC_{LP,SPL}=16.3$ and $\Delta AIC_{LP,SPL} =10.9$ respectively.  This indicates that for these two objects a larger time period than was available for the 2FGL was required to identify curvature in the spectra.  Of the $20$ objects which were reported in the 2FGL as having LP spectra, $10$ had values for $\beta$ that were consistent at the $1 \sigma$ level with the values found using longer observation times in this work.  The values in the 2FGL that were not consistent scattered equally above and below the values found in this work.  This suggests that once enough data have been collected for a source to identify curvature in the spectrum this value does not then change simply by collecting more data, e.g. the spectra do not appear as having more curvature as more data is collected.

\subsection{The Effect of the Instrument Response Function}
Amongst the objects in the clean sample are $3$ FSRQs that were studied in Paper~I and found there to be described significantly better by a BPL than an LP, in agreement with \citet{Abdo10}, but are described significantly better by an LP in this work. There are $3$ possible explanations for this discrepancy: a) by including more data the time-averaged spectra become log-parabolic, b) the difference is due to using an unbinned analysis in Paper~I and a binned analysis in this work or c) the difference is due to using the P6\_V3\_DIFFUSE IRF in Paper~I and the newer P7SOURCE\_V6 IRF in this work.  To determine which explanation was more likely, we ran binned and unbinned analyses on the $3$ objects using the time and energy cuts in Paper~I using the P6\_V3\_DIFFUSE, P6\_V11\_DIFFUSE and P7SOURCE\_V6 IRFs and the appropriate and isotropic and Galactic emission models.  We found that the decision to use a binned or unbinned analysis did not affect the result but that analyses using both the P6\_V3\_DIFFUSE and P6\_V11\_DIFFUSE IRFs identified all $3$ sources as having BPL descriptions while using P7SOURCE\_V6 identified the sources having LP descriptions.  It is clear from this that the differences between the Pass6 and Pass7 IRFs (see Ackermann et al. 2012) have a significant effect in determining the spectrum of an object.  

We then ran an unbinned analysis on all of the objects from Paper~I using the time and energy cuts from that work and using the P6\_V11\_DIFFUSE and P7SOURCE\_V6 IRFs.\footnote{Extending this comparison to a wider range of objects is problematic.  P6\_V3 data are only available to MJD~55707 which is only enough time to distinguish detailed spectral shapes if an object is very bright.}  For each object, the value for $\Delta AIC_{LP,BPL}$ found using P6\_V3\_DIFFUSE is shown in Table~\ref{IRFTable} along with the change in this value using P6\_V11\_DIFFUSE and P7SOURCE\_V6.  As can be seen, using P6\_V11\_DIFFUSE, the $AIC$ value for most objects finds a BPL  to be a better description compared to the LP than in P6\_V3\_DIFFUSE, although some objects still favour an LP.  Going from P6\_V3\_DIFFUSE to P7SOURCE\_V6, some objects favour an LP more strongly and some a BPL more strongly, with those favouring an LP more strongly generally seeing a larger effect.

\begin{figure}
\includegraphics[scale=0.55]{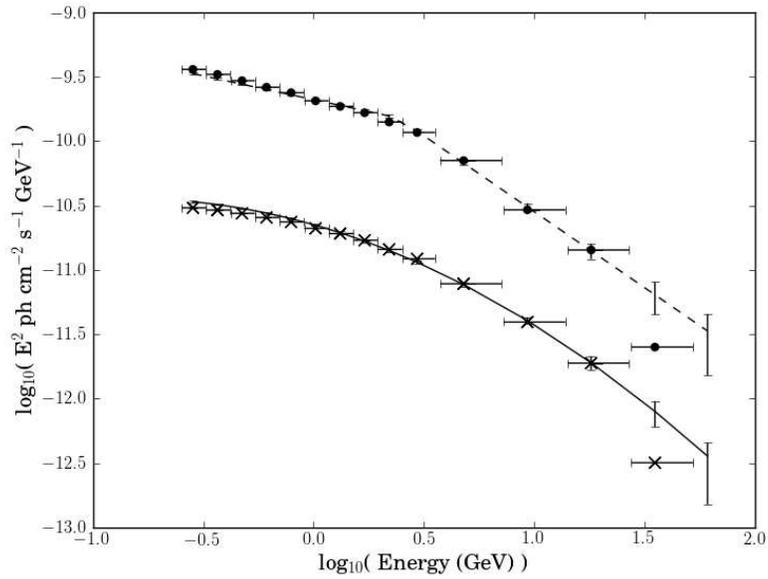}
																			   
\caption{\label{fig:AppPhot}Aperture photometry results from 2FGLJ2253.9+1609 (3C~454.3) for illustrative purposes.  The data points are taken using small, $1^{\circ}$ regions of interest around the source, relying on this to keep the signal to noise high rather than modelling the background.  Circles are results obtained using P6\_V3\_DIFFUSE IRF and crosses are results obtained using P7SOURCE\_V6.  In each case the best fitting model is displayed along side the data.  Each model has 68\% confidence level error bars for the energies of each data point (see \citet{aggarwal2012}).  The P7 data and model have been divided by an arbitrary factor of 10 for clarity.}

\end{figure}

Although there does not appear to be a systematic bias towards one shape or the other, updating the IRF has a strong effect on which spectral shape is favoured.  For illustrative purposes, the results of the aperture photometry analysis of 2FGLJ2253.9+1609 (3C~454.3) over the time period used in Paper~I, analysed using Pass6 and Pass7 IRFs, along with the best models from the unbinned analyses, are shown in Figure~\ref{fig:AppPhot}.  As stated previously, using the most recent IRF, Pass7, an LP is an adequate description for all objects in the clean sample, and in most cases is significantly preferred over a BPL.

\subsection{Light curve analysis}

\begin{figure}
\includegraphics[scale=0.35]{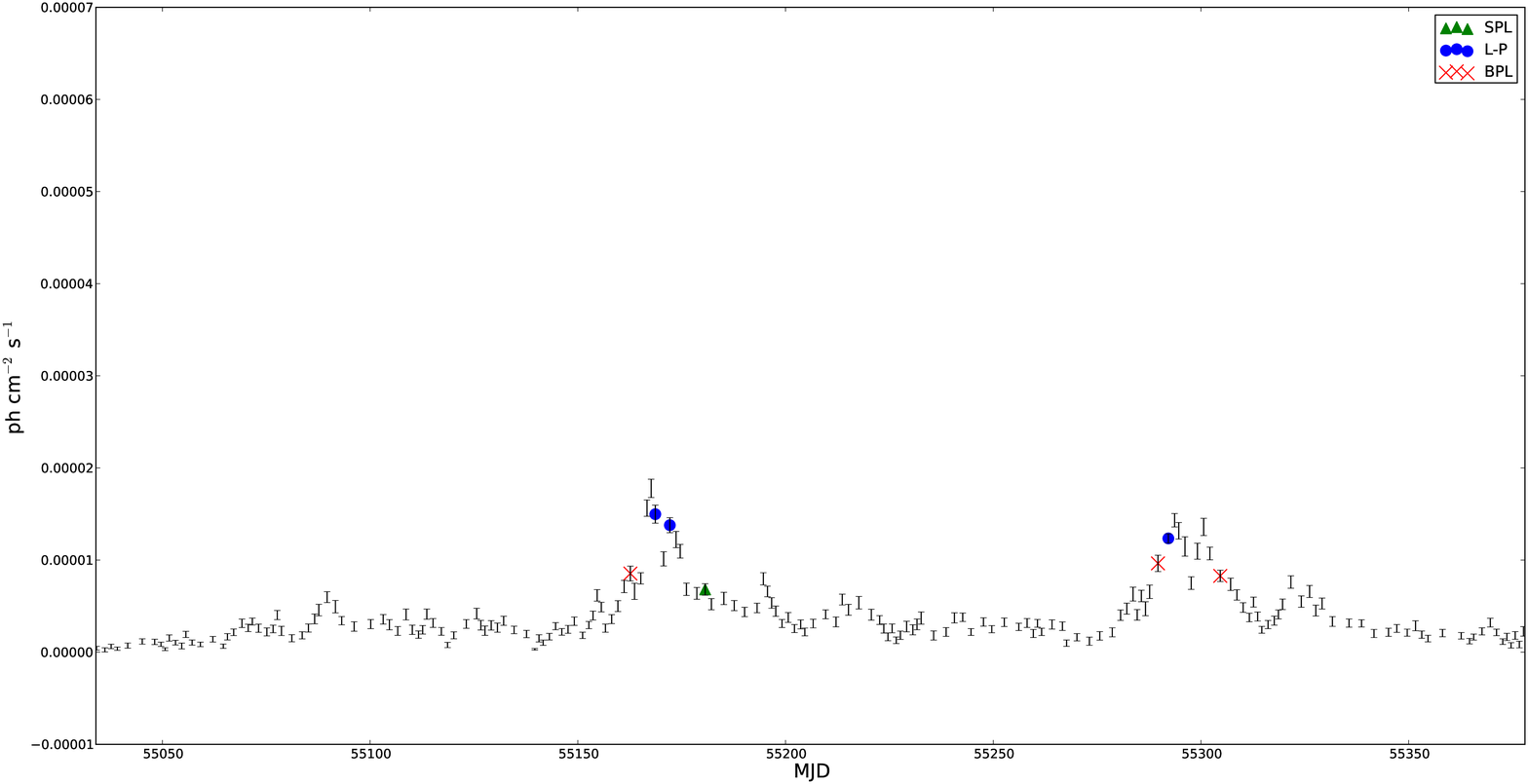}
\includegraphics[scale=0.35]{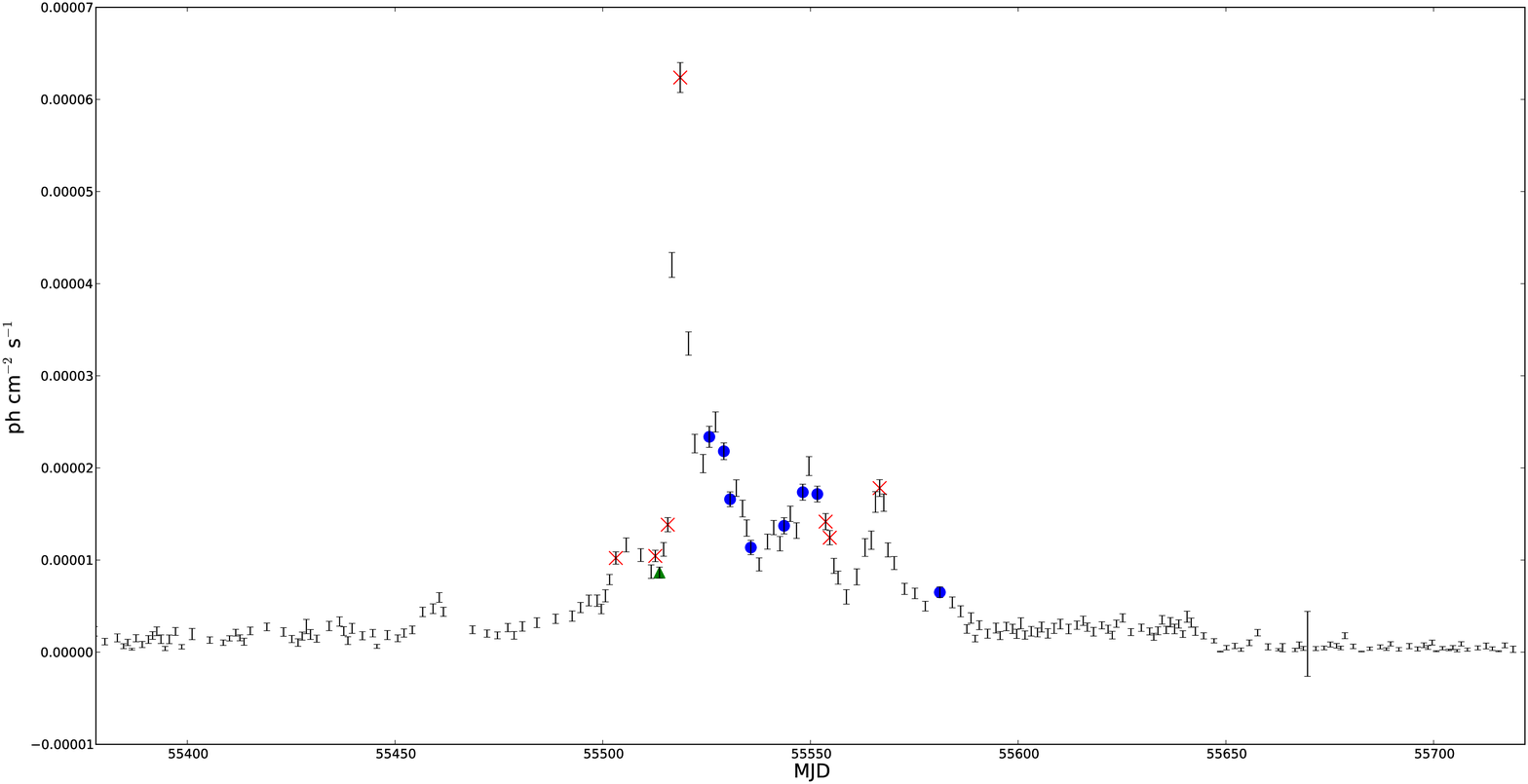}
\caption{\label{fig:3c454LC}Part of the light curve of 2FGLJ2253.9+1609 (3C~454.3).  The symbols indicate Bayesian blocks which are significantly better described by a simple power law (triangles), broken power law (crosses) and log-parabola (circles) models.  Blocks with no symbol saw no significant deviation between models.  The high states between $~$MJD 55150 and $~$MJD 55300 are best described by log-parabolas at their peaks, but the high state at $~$MJD 55500 is best described by a broken power law at its peak.  The time periods either side of that shown do not have any blocks with significant deviation between models.  A colour version of this figure is available in the online version of this article.}

\end{figure}

As shown in Figure~\ref{fig:3c454LC}, the light curve of J2253.9+1609 has three prominent states of high flux.  The spectral behaviour in these three states does not appear to be consistent.  The high states around MJD 55150 and MJD 55300 were significantly better described by LPs at their peaks, with BPL blocks on either side of the peaks.  However, the high state around MJD 55530 was significantly better described by a BPL at its peak, with a roughly even number of BPL and LP blocks on either side.  In total, only two blocks were described significantly better by an SPL.  These results are consistent with the work of \citet{Brown13} and \citet{Rani13} who found respectively that flares in 2FGLJ1512.8-0906 (PKS~1510-08) and 2FGLJ1229.1+0202 (3C~273) were sometimes best described by a BPL and sometimes best described by an LP.

By analysing together the data from all of the quiescent blocks it was found that the quiescent state was significantly better described by an LP than a BPL, with $\Delta AIC_{LP,BPL} = 19$. The LP fit gave $k=8.6 \pm 0.1 \times 10^{-11}~\rm{ph}~\rm{cm}^{-2}~\rm{s}^{-1}$, $\Gamma=2.59\pm 0.01$ and $\beta=0.133\pm 0.007$ giving $E_{P} = 109 \pm 14\rm~MeV$.  Comparing these values with those in Table~\ref{LPTable} we can see that the quiescent and time-averaged values for $\beta$ were consistent, while $\Gamma$ was slightly harder when the higher flux states were included.  We then compared the values of $\Gamma$, $\beta$, and $E_{P}$ in the individual blocks of quiescent flux to the values found analysing all of these blocks together.  At the $1 \sigma$ level the value of $\Gamma$ agreed in $59\%$ of blocks, the value of $\beta$ agreed in $64\%$ of blocks, and the value of $E_{P}$ agreed in $63\%$ of blocks.  This suggests that when in its quiescent state J2253.9+1609 has a fairly stable spectrum.

We now move on to ask what could cause the observed increases in the flux from the quiescent state.  As shown by \citet{Bottcher99}, changes in the mean energy or energy density of an external photon field do not lead to changes in the bolometric Compton luminosity.  However, such changes in the external photon field do cause the Compton peak energy to change and this could cause the luminosity that \textit{Fermi} measures to increase if the peak shifted further into \textit{Fermi}'s energy range.  Because there is a large difference between the quiescent value of $k$ and the time-averaged value this scenario seems unlikely and changes in the external photon field are therefore disfavoured as the cause of the observed high flux states in the light curve.  In an SSC scenario, an increase in the source's  magnetic field could drive an increase in flux and would also increase the value of $\beta$ \citep{Tramacere11}.  This scenario is also disfavoured by our results since $\beta$ is very consistent between its quiescent value and its time-averaged value.

Additional information can be drawn from the value of the Compton peak energy, $E_{P}$.  The LP fit to the quiescent state had $E_{P} = 109 \pm 14\rm~MeV$, which is lower than the time-averaged value of $E_{P}=165 \pm 16 \rm~MeV$ at the $2.6\rm\sigma$ significance level.  If $E_{P}$ does indeed increase along with flux then this could be explained by an increase in the Lorentz factor of the emission region (in an SSC scenario) or a change in the population of the electrons in the emission region (in either an EC or an SSC scenario), such as an increase in the acceleration rate.  On the other hand, if the difference is purely statistical and $E_{P}$ remains constant as the flux increases then this could be explained if the emission was EC in origin and flux increases were driven by increases in the Lorentz factor.

\begin{figure*}
\includegraphics[scale=0.45]{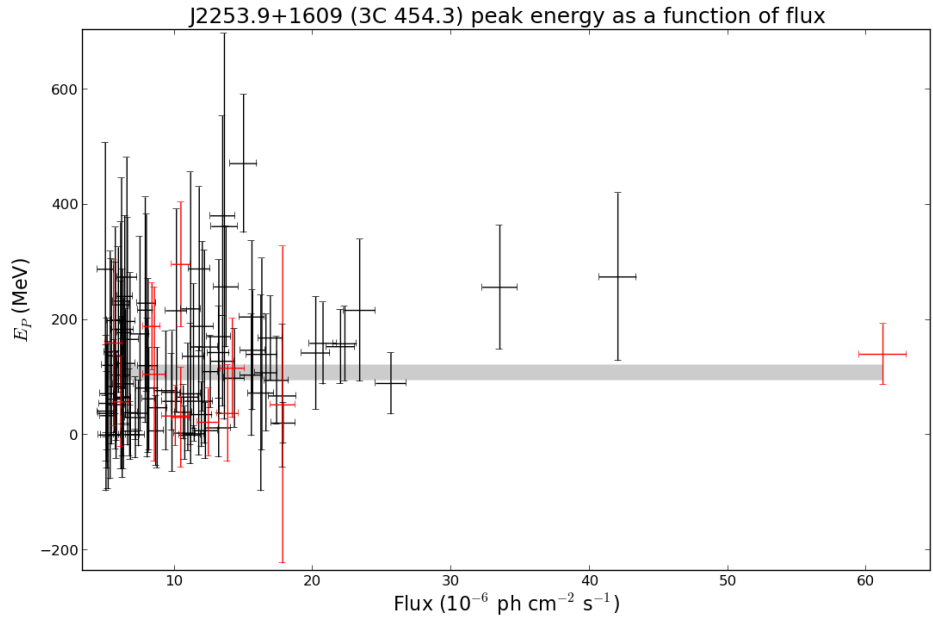}

\caption{\label{fig:peak_ens}Peak energy as a function of flux for 2FGLJ2253.9+1609 (3C~454.3).  Data points are shown for all blocks in a high flux state.  The shaded band shows the 68\% confidence interval for the peak energy in the quiescent state.  Blocks best described by a BPL are shown in red.}

\end{figure*}

To investigate further, we then determined $E_{P}$ for each block in a high flux state that was consistent with an LP spectrum and compared these values to the quiescent value.  The results are shown in Figure~\ref{fig:peak_ens}.  No trend in the value for $E_{P}$ with flux was apparent in the flaring blocks however there are appreciable uncertainties in the data points.  Similarly, light curve analyses of the next brightest objects, 2FGLJ1512.8-0906 (PKS 1510-089), J1229.1+0202 (3C~273), and J1224.9+2122 (4C~21.35) showed no trend between $E_{P}$ and flux, nor was there any trend apparent in the value of $\beta$ with flux.  As more high flux states are observed over time it may become possible to determine their cause.

The BPL blocks in the light curve of J2253.9+1609 are very intriguing. We examined all of the $14$ blocks that were best described by BPL spectra to see if this might be due to the superposition of an additional component on top of the one responsible for the quiescent flux.  For the quiescent component we used the parameters found for analysing all of the quiescent blocks together. We modelled the additional component as an SPL with free parameters and as an LP with free parameters.  In both cases, the difference in $AIC$ showed the BPL model to be a significantly better description than a model with an additional component, so this hypothesis is rejected.

Given that these blocks are well described by a BPL it is natural to try and explain the cause of the break.  To this end  we investigated whether the break energies for blocks best described by a BPL were consistent with the double-absorber model \citep{Poutanen10}.  This model predicts a break at $4.8\rm~GeV$ in an object's rest frame spectrum due to pair absorption of $\gamma $-rays above this energy with low-energy photons from Helium recombination.  Such a result would indicate the emission region was within the radius of the BLR.  Since the double-absorber model also predicts a second break at higher energies which would affect fitting a BPL to the spectrum, to test for a $4.8\rm~GeV$ break it was necessary to make a high energy cut at $6.5\rm~GeV$ in the observer frame, as described in Paper~I. The break energies for each block, along with the exclusion confidence of a break at $4.8\rm~GeV$, are shown in Table~\ref{DATable}. In $12$ of the $14$ blocks which favoured a BPL spectrum, the break energy predicted by the double-absorber model was excluded with $50\%$ confidence or greater. Although it is worth noting that in $1$ case the break appeared where predicted by the double-absorber model, given the total number of trials there is no evidence to favour the double-absorber model and the emission region can be constrained as outside the radius of the BLR.  The model of \citet{Finke10} also seemed unable to explain the appearance of BPL blocks, since this model predicted an approximately stable break energy and, as can be seen in Table~\ref{DATable}, the observed break energies display a considerable variation.

\section{Conclusions}
In this work we analysed $15$~BL Lacs and $27$~FSRQs for evidence of curvature or breaks in their spectra.  We concluded that a log-parabola (LP) is generally the best description of the spectrum.  Objects that seem well-described by a broken power law (BPL) often have relatively bright sources nearby, and the spectral fit seems to be a result of source confusion.  The fact that several objects have previously been identified as having BPL spectra may be an artifact of the older Pass6 IRFs.  It seems rare for the spectrum of a blazar to be best described by a simple power law (SPL); we find only one such object, 2FGLJ0957.7+5522 (4C~+55.17).

Light curve analysis showed that 2FGLJ2253.9+1609 (3C~454.3) appears to have a fairly stable spectrum and is well-described by an LP.  By comparing the parameters found by fitting an LP to the quiescent state with those found by fitting to the whole dataset we can constrain the cause of the flux increases.  Changes in an external photon field (in an EC scenario) and changes in the magnetic field (in an SSC scenario) are disfavoured as the cause of the flux increases.  The cause could be a change in the electron population or Lorentz factor of the emission region.  It may well become possible to distinguish between these two cases as more data are collected.

During high flux states, deviation from an SPL continued, with roughly a dozen blocks each being best-described by an LP and a BPL but only two best-described by an SPL. The appearance of high flux states best-described by a BPL is intriguing and difficult to explain.  The energy of the breaks generally excluded the double-absorber model of \citet{Poutanen10} to a high confidence, suggesting that the emission region was outside the radius of the BLR.  The model of \citet{Finke10} did not seem able to explain the observed BPL spectra either, since this model predicted an approximately stable break energy whilst the break energies showed large variation. 

\section{Acknowledgements}JH acknowledges funding from the UK STFC under grant ST/I505656/1.  We would like to thank the \textit{Fermi} team for the public provision of data and their help desk.  We thank Shangkari Balenderan and Thomas Armstrong for their careful reading of the paper.

\begin{table*}
 \centering
 \begin{minipage}{140mm}
  \caption{AIC results for the clean sample \label{AICTable}}
  \begin{tabular}{@{}lrrr@{}}
  \hline
   Object Name & $\Delta AIC_{min,SPL}$ & $\Delta AIC_{min,BPL}$ & $\Delta AIC_{min,LP}$ \\
  \hline
 BL Lacs & & \\
J0428.6-3756 & 13.5 & 0.6 & 0.0 \\
J0538.8-4405 & 86.8 & 2.9 & 0.0 \\
J0721.9+7120 & 35.4 & 2.8 & 0.0 \\
J2158.8-3013 & 14.7 & 1.8 & 0.0 \\
J2202.8+4216 & 51.0 & 2.3 & 0.0 \\
\hline
 FSRQs & & \\
J0136.9+4751 & 19.9 & 1.7 & 0.0 \\
J0403.9-3604 & 138.7 & 13.6 & 0.0 \\
J0457.0-2325 & 248.6 & 22.8 & 0.0 \\
J0730.2-1141 & 24.9 & 4.6 & 0.0 \\
J0920.9+4441 & 25.4 & 0.0 & 1.7 \\
J0957.7+5522 & 0.0 & 3.0 & 1.8 \\
J1159.5+2914 & 40.3 & 2.7 & 0.0 \\
J1224.9+2122 & 82.8 & 10.1 & 0.0 \\
J1229.1+0202 & 25.7 & 0.0 & 0.1 \\
J1256.1-0547 & 85.8 & 5.5 & 0.0 \\
J1457.4-3540 & 20.7 & 3.5 & 0.0 \\
J1512.8-0906 & 213.7 & 13.3 & 0.0 \\
J1522.1+3144 & 109.5 & 4.0 & 0.0 \\
J1849.4+6706 & 39.9 & 6.0 & 0.0 \\
J2253.9+1609 & 487.9 & 37.5 & 0.0 \\
\hline
 \end{tabular}
 \\$\Delta AIC_{min,s}$ is the difference in $AIC$ value between a model $s$ and the best model for that source.  A value of $0$ indicates the best description of the data and values $>2$ indicate a description is significantly worse than one of the others tested.
 \end{minipage}
 \end{table*}

\begin{table*}
 \centering
 \begin{minipage}{140mm}
  \caption{AIC results for the unclean sample \label{AICTableContam}}
  \begin{tabular}{@{}lrrr@{}}
  \hline
   Object Name & $\Delta AIC_{min,SPL}$ & $\Delta AIC_{min,BPL}$ & $\Delta AIC_{min,LP}$ \\
  \hline
 BL Lacs & & \\
J0222.6+4302 & 3.4 & 0.0 & 5.4 \\
J0238.7+1637 & 22.7 & 2.6 & 0.0 \\
J0449.4-4350 & 2.9 & 0.0 & 0.2 \\
J0818.2+4223 & 23.4 & 3.6 & 0.0 \\
J1015.1+4925 & 15.6 & 1.1 & 0.0 \\
J1104.4+3812 & 9.2 & 0.6 & 0.0 \\
J1427.0+2347 & 15.7 & 2.1 & 0.0 \\
J1542.9+6129 & 0.9 & 1.4 & 0.0 \\
J1555.7+1111 & 11.5 & 0.0 & 0.3 \\
J1653.9+3945 & 9.8 & 0.0 & 4.8 \\
\hline
 FSRQs & & \\
J0108.6+0135 & 39.5 & 5.7 & 0.0 \\
J0442.7-0017 & 4.6 & 1.4 & 0.0 \\
J0719.3+3306 & 8.3 & 0.0 & 0.1 \\
J0725.3+1426 & 9.8 & 1.8 & 0.0 \\
J0808.2-0750 & 16.9 & 0.0 & 2.0 \\
J1246.7-2546 & 91.5 & 10.4 & 0.0 \\
J1312.8+4828 & 7.8 & 3.0 & 0.0 \\
J1428.0-4206 & 73.6 & 7.7 & 0.0 \\
J1504.3+1029 & 94.7 & 8.2 & 0.0 \\
J1635.2+3810 & 112.8 & 14.8 & 0.0 \\
J2025.6-0736 & 71.6 & 2.0 & 0.0 \\
J2056.2-4715 & 55.4 & 1.1 & 0.0 \\
\hline
 \end{tabular}
 \\$\Delta AIC_{min,s}$ is the difference in $AIC$ value between a model $s$ and the best model for that source.  A value of $0$ indicates the best description of the data and values $>2$ indicate a description is significantly worse than one of the others tested.
 \end{minipage}
 \end{table*}

\begin{table*}
 \centering
 \begin{minipage}{140mm}
  \caption{Values for log-parabola fits to the clean sample in the energy range $100\rm~MeV$ to $100\rm~GeV$\label{LPTable}}
  \begin{tabular}{@{}lrrrr@{}}
  \hline
   Object Name &  \multicolumn{1}{c}{Normalisation} &  \multicolumn{1}{c}{Spectral} &  \multicolumn{1}{c}{Curvature} \\
               &    \multicolumn{1}{c}{Constant}    &    \multicolumn{1}{c}{Index}  &  \multicolumn{1}{c}{Parameter}\\
               &    \multicolumn{1}{c}{$10^{-11}\rm~ph~cm^{-2}~s^{-1}~MeV^{-1}$}          &  \multicolumn{1}{c}{$\Gamma$}&  \multicolumn{1}{c}{$\beta$} \\
  \hline
 BL Lacs & & & \\
J0428.6-3756 & 2.1 $\pm$ 0.2 & 1.7 $\pm$ 0.1 & 0.16 $\pm$ 0.04 \\
J0538.8-4405 & 3.98 $\pm$ 0.05 & 2.06 $\pm$ 0.01 & 0.069 $\pm$ 0.008 \\
J0721.9+7120 & 2.50 $\pm$ 0.04 & 1.97 $\pm$ 0.03 & 0.09 $\pm$ 0.01 \\
J2158.8-3013 & 1.97 $\pm$ 0.04 & 1.82 $\pm$ 0.01 & 0.027 $\pm$ 0.007 \\
J2202.8+4216 & 2.16 $\pm$ 0.04 & 2.26 $\pm$ 0.01 & 0.061 $\pm$ 0.009 \\
\hline
 FSRQs & & & \\
J0136.9+4751 & 0.74 $\pm$ 0.03 & 2.34 $\pm$ 0.03 & 0.08 $\pm$ 0.02 \\
J0403.9-3604 & 1.14 $\pm$ 0.03 & 2.69 $\pm$ 0.03 & 0.18 $\pm$ 0.02 \\
J0457.0-2325 & 2.71 $\pm$ 0.04 & 2.27 $\pm$ 0.01 & 0.126 $\pm$ 0.009 \\
J0730.2-1141 & 2.52 $\pm$ 0.05 & 2.35 $\pm$ 0.02 & 0.06 $\pm$ 0.01 \\
J0920.9+4441 & 0.92 $\pm$ 0.03 & 2.37 $\pm$ 0.03 & 0.10 $\pm$ 0.02 \\
J0957.7+5522 & 1.4 $\pm$ 0.9 & 2.1 $\pm$ 0.6 & 0.1 $\pm$ 0.1 \\
J1159.5+2914 & 1.10 $\pm$ 0.03 & 2.27 $\pm$ 0.02 & 0.08 $\pm$ 0.01 \\
J1224.9+2122 & 3.60 $\pm$ 0.05 & 2.34 $\pm$ 0.01 & 0.054 $\pm$ 0.006 \\
J1229.1+0202 & 1.82 $\pm$ 0.04 & 2.82 $\pm$ 0.02 & 0.09 $\pm$ 0.02 \\
J1256.1-0547 & 3.14 $\pm$ 0.05 & 2.42 $\pm$ 0.01 & 0.066 $\pm$ 0.008 \\
J1457.4-3540 & 1.04 $\pm$ 0.04 & 2.38 $\pm$ 0.03 & 0.10 $\pm$ 0.02 \\
J1512.8-0906 & 6.49 $\pm$ 0.06 & 2.447 $\pm$ 0.009 & 0.070 $\pm$ 0.005 \\
J1522.1+3144 & 2.82 $\pm$ 0.04 & 2.42 $\pm$ 0.01 & 0.074 $\pm$ 0.008 \\
J1849.4+6706 & 0.79 $\pm$ 0.03 & 2.20 $\pm$ 0.03 & 0.11 $\pm$ 0.02 \\
J2253.9+1609 & 17.5 $\pm$ 0.1 & 2.475 $\pm$ 0.007 & 0.132 $\pm$ 0.007 \\
\hline
 \end{tabular}
 \end{minipage}
 \end{table*}

\begin{table*}
 \centering
 \begin{minipage}{140mm}
  \caption{Comparison of spectral fits using different instrument response functions.\label{IRFTable}}
  \begin{tabular}{@{}lrrr@{}}
  \hline
   Object Name & \multicolumn{1}{c}{$\Delta AIC_{LP,BPL}$} &\multicolumn{1}{c}{Change when} &\multicolumn{1}{c}{Change when}  \\
    & \multicolumn{1}{c}{using $P6\_V3$} &\multicolumn{1}{c}{using $P6\_V11$} &\multicolumn{1}{c}{using $P7\_V6$}  \\   
  \hline
J0457.0-2325 & 1.8 & 3.6 & 5.6 \\
J0920.9+4441 & 13.2 & -11.1 & -1.3\\
J1229.1+0202 & -1.5 & -3.9 & 18.0 \\
J1256.1-0547 & 6.6 & -7.7 & -3.5 \\
J1504.3+1029 & 2.6 & -6.0 & -0.3 \\
J1512.8-0906 & -2.4 & -2.6 & 4.9 \\
J1522.1+3144 & -9.2 & -2.6 & 13.9 \\
J2025.6-0736 & -0.4 & 0.5 & 2.3 \\
J2253.9+1609 & -12.8 & -16.0 & 65.7 \\

\hline

 \end{tabular}
 \\The second column gives the difference in $AIC$ values when fitting the spectrum with a Broken Power Law and a Log-Parabola using the P6\_V3 IRF.  A positive value shows that LP was favoured and a negative value indicates a BPL was favoured.  The third and fourth columns show how this value changes using the P6\_V11 and P7\_V6 IRFs respectively.  A positive value shows that an LP fit is better than when using P6\_V3 IRF and a negative value shows that a BPL fit is better than when using P6\_V3 IRF.
 \end{minipage}
 \end{table*}
 
\begin{table*}
 \centering
 \begin{minipage}{140mm}
  \caption{Break energies for blocks in the light curve of 2FGLJ2253.9+1609 (3C~454.3) that were best described by a broken power law, and the exclusion confidence of a $4.8$\rm~GeV break, as predicted by the double-absorber model \citep{Poutanen10}.\label{DATable}}
  \begin{tabular}{@{}cccc@{}}
  \hline
   \multicolumn{1}{c}{Block start} & \multicolumn{1}{c}{Block Length} &\multicolumn{1}{c}{Break Energy} & \multicolumn{1}{c}{4.8 GeV Exclusion} \\
   \multicolumn{1}{c}{(MJD)} & \multicolumn{1}{c}{(Days)} &\multicolumn{1}{c}{(GeV)} & \multicolumn{1}{c}{Confidence} \\   
  \hline
55089 & 3 & $4.1^{+0.7}_{-0.3}$ & 65\% \\
55158 & 2 & $4.1^{+1.2}_{-0.6}$ & 59\% \\
55163 & 1 & $4^{+2}_{-2}$ & 53\% \\
55215 & 2 & $3^{+1}_{-1}$ & 76\% \\
55289 & 3 & $0.4^{+0.2}_{-0.2}$ & 84\% \\
55304 & 3 & $1.2^{+0.2}_{-0.3}$ & $>$99\% \\
55459 & 2 & $7^{+2}_{-1}$ & 79\% \\
55499 & 1 & $3.2^{+0.7}_{-1.3}$ & 93\% \\
55503 & 1 & $2.8^{+0.6}_{-0.7}$ & 83\% \\
55512 & 1 & $5^{+\infty}_{-\infty}$ & $<$1\% \\
55516 & 2 & $5.2^{+1.0}_{-0.8}$ & 31\% \\
55518 & 1 & $2.1^{+0.7}_{-0.8}$ & 97\% \\
55555 & 3 & $1.0^{+0.3}_{-0.2}$ & $>$99\% \\
55567 & 1 & $0.6^{+0.2}_{-0.2}$ & 86\% \\
\hline

 \end{tabular}
 \end{minipage}
 \end{table*}

 

\clearpage

\clearpage

\clearpage

\clearpage

\label{lastpage}

\end{document}